# Two-dimensional chemical mapping for non-coding RNAs

Wipapat Kladwang[1], Christopher C. VanLang[2], Pablo Cordero[3], and Rhiju Das[1,3,4]*

*Departments of Biochemistry[1] & Chemical Engineering[2], Program in Biomedical Informatics[3], and Department of Physics[4], Stanford University, Stanford CA 94305*

- *To whom correspondence should be addressed: rhiju@stanford.edu. Phone: (650) 723-5976. Fax: (650) 723-6783.*

**Abstract**

Non-coding RNA molecules fold into precise base pairing patterns to carry out critical roles in genetic regulation and protein synthesis. We show here that coupling systematic mutagenesis with high-throughput SHAPE chemical mapping enables accurate base pair inference of domains from ribosomal RNA, ribozymes, and riboswitches. For a six-RNA benchmark that challenged prior chemical/computational methods, this mutate-and-map strategy gives secondary structures in agreement with crystallographic data (2% error rates), including a blind test on a double-glycine riboswitch. Through modeling of partially ordered RNA states, the method enables the first test of an 'interdomain helix-swap' hypothesis for ligand-binding cooperativity in a glycine riboswitch. Finally, the mutate-and-map data report on tertiary contacts within non-coding RNAs; coupled with the Rosetta/FARFAR algorithm, these data give nucleotide-resolution three-dimensional models (5.7 Å helix RMSD) of an adenine riboswitch. These results highlight the promise of a two-dimensional chemical strategy for inferring the secondary and tertiary structures that underlie non-coding RNA behavior.

The transcriptomes of living cells and viruses continue to reveal novel classes of non-coding RNA (ncRNA) with critical functions in gene regulation, metabolism, and pathogenesis (see, e.g., refs[1-7]). The functional behaviors of these molecules are intimately tied to specific base-pairing patterns that are challenging to identify by existing strategies based on phylogenetic analysis, NMR[8-11], crystallography[12-17], molecular rulers[18, 19], or functional mutation/rescue experiments (see, e.g., refs[7, 20, 21]). A more facile approach to characterizing RNA structure involves high-throughput chemical mapping at single-nucleotide resolution. This method is applicable to RNAs as large as the ribosome or entire viruses both *in vitro* and in their cellular milieu.[22-25] Measurements of every nucleotide's accessibility to solution chemical modification can guide or filter structural hypotheses from computational models[26-28]. Nevertheless, approximations in computational models and in correlating structure to chemical accessibility limit the inherent accuracy of this approach.[26-31]

This paper presents a strategy to expand the information content of chemical mapping through a two-dimensional 'mutate-and-map' methodology.[32] Here, sequence mutation acts as a second dimension in a manner analogous to initial perturbation steps in multidimensional NMR pulse sequences[11] or pump/probe experiments in other spectroscopic fields[33]. We reasoned that if one nucleotide involved in a base pair is mutated, its partner might become more exposed and thus be readily detectable by chemical mapping. In practice, some mutations might not lead to the desired 'release' of the pairing partners; and some mutations might produce larger perturbations, such as the unfolding of an entire helix. Nevertheless, if even a subset of the probed mutations leads to precise release of interacting nucleotides, the RNA's base pairing pattern could potentially be read out from this extensive data set. Indeed, our recent proof-of-concept studies demonstrated the systematic inference of Watson/Crick base pairs in a 20-base-pair DNA/RNA duplex[32] and a 35-nucleotide RNA hairpin[34]. Nevertheless, these artificial systems were designed to include single long helices and thus may not adequately represent natural, functional non-coding RNAs with many shorter helices, extensive non-canonical interactions, and multiple solution states.

We therefore sought to apply the mutate-and-map strategy to a diverse set of non-coding RNAs with available crystal structures for some states and unknown structures for other states. The benchmark, which includes ribozymes, riboswitches, and ribosomal RNA domains (Table S1), is challenging: a prior (one-dimensional) chemical/computational approach missed and mispredicted ~20% of the benchmark's helices.[29] We have found that the mutate-and-map strategy achieves 98% accuracy in inferring Watson/Crick base pairing patterns and gives clear confidence estimates through bootstrap analysis. Furthermore, the method permits the generation and falsification of structural hypotheses about partially ordered RNA states, as highlighted by new results on a glycine-sensing riboswitch. Our main focus herein is on the basic but unsolved problem of RNA secondary structure inference[35-37] from biochemical data. Extensions of the method and advances in computational modeling may permit robust tertiary contact inference and 3D models, and we present one such case as a proof-of-concept.

## Results

*Proof-of-concept on an adenine-binding riboswitch*

We first established the information content and accuracy of the strategy on the 71-nucleotide adenine-sensing *add* riboswitch from *V. vulnificus*, which has been extensively studied (see, e.g,. refs[9, 20, 38-40]) and solved in the adenine-bound state by crystallography[13]. The *add* secondary structure is incorrectly modeled by the *RNAstructure* algorithm alone but can be recovered through the inclusion of standard 1D SHAPE (2´-OH acylation) data[29]; this RNA therefore serves as a well-characterized control. We prepared 71 variants of the RNA, mutating each base to its complement, using high-throughput PCR assembly, *in vitro* transcription, and magnetic bead purification methods in 96-well format, as discussed previously[34]. Figure 1A gives the SHAPE electropherograms for the wild type construct and for each mutant in the presence of 5 mM adenine; the data were collected in a single afternoon. As before, Z-scores (number of standard deviations from the mean accessibility; see Methods) highlight the most significant features of the data.

As with simpler model systems, the *add* mutate-and-map data show distinct diagonal stripes (I in Fig. 1A), corresponding to perturbations at each mutation site, and numerous off-diagonal features corresponding to interacting pairs of sequence-separated nucleotides. For example, the mutation C18G led to strong exposure of G78 and of no other nearby bases (II, Fig. 1A), strongly supporting a C18-G78 base pair in the P1 helix. Such 'punctate' single-nucleotide-resolution base pair features are also visible for the other helical stems of this RNA (e.g., C26-G44 and C54-G72, marked III and IV in Fig. 1A). Several mutations led to more delocalized perturbations (V-VII, Fig. 1A; stems marked in Fig. 1B); while not punctate, these features confirm interactions at lower, helix-level resolution.

To integrate and assess the predictive power of these data, we applied the measured Z-scores as energetic bonuses in the *RNAstructure* secondary structure prediction algorithm. We further estimated confidence values for all inferred helices by bootstrapping the mutate-and-map data[41] and repeating the secondary structure calculation. As expected from the visual analysis above, the crystallographic secondary structure was robustly recovered, with $\geq$99% bootstrap values for the helices P1, P2, and P3. An "extra" 2-bp helix was also found, with a weak bootstrap value (58%); these nucleotides are in fact base-paired in the *add* riboswitch crystallographic model[13] but one pairing is a non-canonical Watson-Crick/Hoogsteen pair and part of a base triple. Along with additional mutate-and-map signals (VII–X), these data were sufficient for determining the RNA's global tertiary fold, as is described below.

*A challenging benchmark of base pair inference*

To complete our benchmark of the mutate-and-map strategy, we applied the method to RNAs whose base pairing patterns have been more challenging to recover. The smallest of these, unmodified tRNA$^{phe}$ from *E. coli*[14], offers a simple illustration of the new method's information content (Fig. 2A). The *RNAstructure* algorithm mispredicted two of the four helices of the tRNA 'cloverleaf' (the D and the anticodon helices; Fig. 2C). Inclusion of 1D SHAPE data corrected these errors, but introduced an additional error, mispredicting the T$\psi$C helix (Fig. 2E). The mutate-and-map SHAPE data for this tRNA (Fig. 2A) gave clear signals for all four helices. Applying these 2D data to *RNAstructure* corrected the algorithm's inherent inaccuracies and recovered the entire four-helix secondary structure (>99% bootstrap values; Fig. 2D). One additional edge base pair was predicted for the anticodon arm; this and other fine-scale errors are discussed below.

The remaining RNAs in our benchmark exceeded 100 nucleotides in length. As in the tRNA$^{phe}$ case, prior chemical/computational methods assigned incorrect secondary structures to these sequences, but the mutate-and-map strategy led to accurate base-pairing patterns.  First, the mutate-and-map data for a widely studied model RNA, the P4-P6 domain of the group I *Tetrahymena* ribozyme[42], gave visible features corresponding to all helices in the RNA[43] (Fig 3A) and led to correct recovery of the secondary structure (Fig. 3B). One of the helices, P5c, was correctly modeled but with a weak bootstrap value (48%); this low score is consistent with conformational fluctuations in P5c identified in previous biochemical and NMR studies[44-46].

As a more stringent test of the mutate-and-map strategy, we applied the method to the *E. coli* 5S ribosomal RNA, a notable problem case for prior chemical/computational approaches[26, 27, 31]. In particular, the segments around the non-canonical loop E motif have been mispredicted in all prior studies, including the most recent (1D) SHAPE-directed approach[29]. By providing pair-wise information on interacting nucleotides (Fig. 2A), the mutate-and-map method recovered the entire secondary structure with high confidence (> 90%; Fig. 2B). One extra helix (blue in Fig. 2B) corresponds to a segment that in fact forms non-canonical base pairs within the loop E motif.

Third, the ligand-binding domain of the cyclic di-GMP riboswitch from *V. cholerae* provided an additional challenge; this RNA's helix P1 was not found in the original phylogenetic analysis[21] but instead later revealed by crystallography. Based on measurements in the presence of 10 μM ligand, the mutate-and-map strategy (Fig. 3E) recovered nearly the entire secondary structure (7 of 8 helices), including P1 (Fig. 3F).

*Blind prediction on the glycine riboswitch*

As a final rigorous test, we acquired mutate-and-map data for an RNA whose crystallographic model was not available at the time of modeling: the ligand-binding domain of the glycine-binding riboswitch from *F. nucleatum*.[47, 48] The mutate-and-map data in the presence of 10 mM glycine gave a secondary structure with 9 helices (Fig. 4A); the model agreed with the 9 helices that were identified by phylogeny. The secondary structure was confirmed by a crystallographic model released at the time of this paper's submission[49].

*Overall accuracy of the mutate-and-map method*

Overall, the mutate-and-map method demonstrated high accuracy in secondary structure inference for a benchmark of six diverse RNAs including 661 nucleotides involved in 42 helices. As a baseline, a prior method, using *RNAstructure* directed by 1D SHAPE data, gave a false negative rate and false discovery rate of 17% and 20%, respectively, on this benchmark.[29] The mutate-and-map method recovered 41 of 42 helices, giving a sensitivity of 98% and a false negative rate of 2%, nearly an order of magnitude less than the prior method. The only missing helix was a two-base-pair helix in the cyclic diGMP riboswitch (see below). Looking at finer resolution, a small number (< 6%) of the base pairs in mutate/map-calculated helices were either missed or added relative to crystallographic secondary structures (1 and 11 of 197 base pairs, respectively; Table S2). All of these errors were either G-U or A-U pairs at the edges of otherwise correct helices (Figs. 1–4 & Table S2).

In terms of the false discovery rate, the mutate-and-map method gave only 3 extra helices, all of which were the smallest possible in length (2 bp). As discussed above, two of these extra helices in fact correspond to non-canonical stems observed in crystallographic models. The remaining false helix gave a weak bootstrap value (60%) and may correspond to a stem sampled in the ligand-free conformation of the cyclic diGMP riboswitch (see below and SI Fig. S3). The overall positive predictive value was 93 to 98% depending on whether the noncanonical helices are counted as correct. The false discovery rate was 2-7%, nearly an order of magnitude less than the prior 1D SHAPE-directed method (20%). Somewhat surprisingly, using both the 1D SHAPE data and 2D mutate-and-map data gave worse accuracy than using the 2D data alone; this result may reflect inaccuracies in interpreting absolute SHAPE reactivity, as opposed to changes in reactivity induced by mutations. We conclude that secondary structures derived from the mutate-and-map method are accurate (~2% error rates) for structured non-coding RNAs.

*Testing an 'inter-domain helix swap' hypothesis for glycine riboswitch cooperativity*

Beyond recovering known information about non-coding RNA secondary structure, we sought to generate or falsify novel hypotheses that would be difficult to explore by standard structural methods. The three riboswitch ligand-binding domains for adenine, cyclic di-GMP, and glycine provide interesting test cases because their ligand-free states will generally be partially ordered and thus difficult to crystallize. First, application of the mutate-and-map strategy indicated that the secondary structure of the *add* riboswitch ligand-binding domain remains the same in adenine-free and adenine-bound states (SI Fig. S2), consistent with biophysical data from other approaches [5, 9, 10]. In contrast, mutate-and-map data indicated that the cyclic di-GMP riboswitch shifts its secondary structure near P1 upon ligand binding (SI Fig. S3). This shift is potentially involved in the riboswitch's mechanism[16, 17, 21] and may account for the weak phylogenetic signature of the P1 helix.

Among these 'non-crystallographic' targets, we were most interested in the glycine-binding riboswitch, which exhibits cooperative binding of two glycines to separate domains and is under intense investigation by several groups. [47-52] Analogous to the tense/relaxed equilbrium in the Monod-Wyman-Changeaux model for hemoglobin[53], we hypothesized that cooperativity might stem from an inter-domain helix swap. In this model, an alternative ('tense') secondary structure involving non-native interactions between the two domains would be rearranged upon glycine binding. The model was not readily testable by prior 1D chemical/computational methods due to their low information content or by crystallography[49, 51], which, if successful, is biased towards more structured conformations.

Application of the mutate-and-map strategy (Figs. 4A & 4C) gave a strong test of the hypothesis: the data in the absence of glycine gave the same domain-separated secondary structure as under conditions with glycine bound (cf. Figs. 4B & 4D). Any changes in secondary structure for these constructs are thus either at edge base pairs or are negligible. We note that additional 5´ and 3´ flanking elements are likely to play critical roles in these RNA's modes of genetic regulation[2]; these longer segments are now under investigation.

*Tertiary structure and cooperative fluctuations*

The analysis described above focused on the first level of RNA structure, the Watson-Crick base-pairing pattern. Nevertheless, many non-coding RNAs use tertiary contacts and ordered junctions to position Watson-Crick helices into intricate three-dimensional structures. Qualitatively, we found evidence for many such tertiary interactions in these RNAs' mutate-and-map data. For example, the *add* riboswitch is stabilized by the tertiary interactions between the loops L2 (nucleotides 32–38) and L3 (nucleotides 60–66). In the presence of 5 mM adenine, mutations at G37 and G38 resulted in exposure of their partners C61 and C60 (VIII, Fig. 1A), and vice versa (IX, Fig. 1A; L2/L3 marked in Fig. 1B). Nevertheless, other mutations led to longer range effects due to cooperative unfolding of subdomains of tertiary structure or loss of adenine binding. For example, mutation of nucleotide A52 gave chemical accessibilities that were different from the adenine-bound wild type RNA throughout the sequence, but appear consistent with the adenine-unbound state.

To effectively discriminate contacts due to secondary or tertiary structure from more disperse effects, we implemented sequence-independent filters enforcing strong, punctate signals and symmetry (see ref[34], Methods and SI Fig. S3). This analysis, independent of any computational models of RNA structure, recovered the majority of Watson-Crick helices in this benchmark. The analysis also recovered three tertiary contacts: a U22-A52 base pair in the adenine binding core and the L2/L3 interaction of the *add* riboswitch; and a tetraloop/receptor interaction in the P4-P6 RNA. These features are accurate, but, in most test cases, their number is significantly less than the number of helices, precluding effective 3D modeling. For the one case in which multiple tertiary contacts could be determined, the *add* adenine-sensing riboswitch, we carried out 3D modeling using the FARFAR *de novo* assembly method. The algorithm gave a structural ensemble with helix RMSD of 5.7 Å and overall RMSD of 7.7 Å to the crystallographic model[13] (Fig. 5A vs. 5B). This resolution is comparable to the average distance between nearest nucleotides (5.9 Å) and significantly better than expected by chance ($P<10^{-3}$ for modeling a 71-nt RNA with secondary structure information[54]). These results on a favorable case suggest that rapid inference of 3D structure for general RNA might be achievable with other chemical probes that discriminate non-canonical interactions (e.g., dimethyl sulfate[34, 55] for A-minor interactions) or more sophisticated methods for mining tertiary information from mutate-and-map data. We note also that features corresponding to cooperative changes in chemical accessibility, while not reporting on specific tertiary contacts, can reveal accessible alternative states in the RNA's folding landscapes. We are making the information-rich data sets acquired for this paper publicly available in the Stanford RNA Mapping Database (http://rmdb.stanford.edu) to encourage the development of novel analysis methods to explore tertiary contact extraction and landscapes.

**Discussion**

We have demonstrated that a mutate-and-map strategy permits the high-throughput inference of non-coding RNA base pairing patterns. With error rates of ~2% and confidence estimates via bootstrapping, the method recapitulates the secondary structures of riboswitch, ribosomal, and ribozyme domains for which prior chemical/computational approaches gave incorrect models. In addition to recovering known structures, the mutate-and-map data permit the rapid generation and falsification of hypotheses for structural rearrangements in three ligand-binding RNAs in partially ordered ligand-free states, including a cooperative glycine riboswitch with a poorly understood mechanism. Finally, the data yield rapid information on tertiary contacts and the energetic architecture of ncRNAs, in one case sufficient to define a riboswitch's 3D helix arrangement at nucleotide resolution (5.7 Å).

The method requires only commercially available reagents, widely accessible capillary electrophoresis sequencers, and freely available software. Further, each data set herein was acquired and analyzed in a week or less. Therefore, for non-coding RNA domains up to ~300 nucleotides in length, the technology should be applicable as a front-line structural tool. Anticipated accelerations based on next-generation sequencing and random barcoding may enable the routine characterization of transcripts with thousands of nucleotides.

Expanding experimental technologies from one to multiple dimensions has transformed fields ranging from NMR to infrared spectroscopy. We propose that the mutate-and-map strategy will be analogously enabling for chemical mapping approaches, permitting the confident secondary structure determination and tertiary contact characterization of non-coding RNAs that are difficult or intractable for previous experimental methods. Applications to full-length RNA messages *in vitro* or in extract, to complex ribonucleoprotein systems, and even to full viral RNA genomes appear feasible and are exciting frontiers for this high-throughput approach.

**Acknowledgements**

We are grateful to the authors of *RNAstructure* for making their source code freely available. This work was supported by the Burroughs-Wellcome Foundation (CASI to RD), NIH (T32 HG000044 to CCV), and a Stanford Graduate Fellowship (to PC).

## Methods

*The mutate-and-map experimental protocol and data processing*

Preparation of DNA templates, *in vitro* transcription of RNAs, SHAPE chemical mapping, and capillary electrophoresis were carried out in 96-well format, accelerated through the use of magnetic bead purification steps, as has been described previously. Data were analyzed with the HiTRACE software package, and Z-scores were computed in MATLAB. A complete protocol is given in Supplementary Methods; code for analyzing mutate-and-map data is being made available as part of HiTRACE. The Z-scores were used for secondary structure inference (see next) and sequence-independent feature analysis by single-linkage clustering, as described in Supplementary Methods.

*Secondary structure inference*

The *Fold* executable of the *RNAstructure* package (v5.1) was used to infer secondary structures. The entire RNA sequences (Table S1), including added flanking sequences, were used for all calculations. The flag "-sh", was used to input (one-dimensional) SHAPE data files. The Z-score data were introduced by modifying the code to take in a text file of base-pair energy bonuses via an additional flag "-x". In the *RNAstructure* implementation, these pseudoenergies are applied to each nucleotide that forms an edge base pair, and doubly applied to each nucleotide that forms an internal base pair, similar to incorporation of 1D SHAPE data[28]. For bootstrap analyses, mock SHAPE data replicates were generated by randomly choosing mutants with replacement [41]. The code modifications are available upon request from the authors; we have also contacted the *RNAstructure* developers to suggest that these modifications be incorporated into a later release.   Secondary structure images were prepared in VARNA [56].

*Assessment of secondary structure accuracy*

A crystallographic helix was considered correctly recovered if more than 50% of its base pairs were observed in a helix by the computational model. (In practice, 39 of 41 such helices in models based on mutate/map data retained all crystallographic base pairs.) Helix slips of ±1 were not considered correct [i.e., the pairing $(i,j)$ was not allowed to match the pairings $(i,j-1)$ or $(i,j+1)$].

*Three-dimensional modeling with Rosetta*

Three-dimensional models were acquired using the Fragment Assembly of RNA with Full Atom Refinement (FARFAR) methodology[57] in the Rosetta framework.

Briefly, ideal A-form helices were created for each helix greater than 2 base-pairs in length in the modeled secondary structure. Then, remaining nucleotides were modeled by FARFAR as separate motifs interconnecting these ideal helices, generating up to 4000 potential structures. Finally, these motif conformations were assembled in a Monte Carlo procedure, optimizing the FARNA low-resolution potential and tertiary constraint potentials defined by the sequence-independent clustering analysis of mutate-and-map data. Explicit command lines and example files are given in Supporting Information. The code, along with a Python job-setup script *create_motif_jobs.py* and documentation, are being incorporated into Rosetta release 3.4, which is freely available to academic users at http://www.rosettacommons.org. Prior to release, the code is available upon request from the authors. The *P*-value for the *add* riboswitch was estimated by comparing the all-atom RMSD (7.7 Å) to the range expected by chance (13.5 Å ± 1.8 Å), as described in [54].

**Table 1.** Accuracy of RNA secondary structure models using one-dimensional SHAPE data and/or two-dimensional mutate/map data.

| RNA | Len. | Number of helices | | | | | | | | |
|---|---|---|---|---|---|---|---|---|---|---|
| | | Cryst. | No data | | 1D SHAPE | | 1D+Mutate/map | | Mutate/map | |
| | | | TP[a] | FP[a] | TP[a] | FP[a] | TP[a] | FP[a] | TP[a] | FP[a] |
| Adenine ribosw.[b] | 71 | 3 | 2 | 3 | 3 | 0 (1) | 3 | 0 (1) | 3 | 0 (1) |
| tRNA[phe] | 76 | 4 | 2 | 3 | 3 | 1 | 4 | 0 | 4 | 0 |
| P4-P6 RNA | 158 | 11 | 10 | 1 | 9 | 2 | 9 | 2 | 11 | 0 |
| 5S rRNA | 118 | 7 | 1 | 9 | 6 | 3 | 7 | 0 (1) | 7 | 0 (1) |
| c-di-GMP ribosw.[b] | 80 | 8 | 6 | 2 | 6 | 2 | 7 | 1 | 7 | 1 |
| Glycine ribosw.[b,*] | 158 | 9 | 5 | 3 | 8 | 1 | 9 | 0 | 9 | 0 |
| *Total* | 661 | 42 | 26 | 21 | 35 | 9 (10) | 39 | 3 (5) | 41 | 1 (3) |
| **False negative rate**[c] | | | 38.1% | | 16.7% | | 7.1% | | 2.4% | |
| **False discovery rate**[d] | | | 44.7% | | 20.4 (22.2)% | | 7.1 (11.4)% | | 2.3 (6.8)% | |

[a] TP = true positive helices; FP = false positive helices. For FP, a helix was considered incorrect if its base pairs did not match the majority of base pairs in a crystallographic helix. Numbers in parentheses required that the matching crystallographic base pairs have Watson-Crick geometry.
[b] Ligand-binding riboswitches were probed in the presence of small-molecule partners (5 mM adenine, 10 μM cyclic di-guanosine-monophosphate, or 10 mM glycine). All experiments were carried out with 10 mM $MgCl_2$, 50 mM Na-HEPES, pH 8.0.
[c] False negative rate = (Total in crystal – TP )/ TP.
[d] False discovery rate = FP/(FP+TP). Numbers in parentheses count matches of model base pairs to non-Watson-Crick crystallographic base pairs as false discoveries.

## Figure legends

**Figure 1. Illustrating the mutate-and-map method on an adenine riboswitch.** **(a)** Chemical mapping (SHAPE) data for 71 single-nucleotide mutants of the *add* adenine-binding domain from *V. vulnificus* in 5 mM adenine. Modifications were read out by high throughput reverse transcription with 5´-fluorescently labeled radiolabeled primers and capillary electrophoresis. Raw fluorescence traces (arbitrary units) are shown after automated alignment and normalization to mean intensity. Shorter products (higher electrophoretic mobility) appear on the right. Ten features are marked on the data: (I) the main diagonal stripe showing localized perturbations upon the C18G mutation; (II-IV) punctate features marking base pairs C18-G78, C26-G44, and C54-G72 in three different helices; (V-VII) more delocalized effects upon helix mutations; (VIII) evidence for long-range tertiary contact between L2 and L3 upon mutation of C60 & C61 in L3; (IX) 'symmetric' mutations in L2 that affect L3; (X) evidence for an U22-A52 base pair in the adenine binding site. **(b)** Z-scores associated with mutate-and-map data. **(c)** Secondary structure derived from incorporating the mutate-and-map Z-scores into the *RNAstructure* modeling algorithm; bootstrap confidence estimates are given in red. Additional tertiary contacts inferred from sequence-independent clustering analysis are given in green.

**Figure 2. Comparison of chemical/computational modeling approaches on $\mathbf{tRNA^{phe}}$.** **(a)** Mutate-and-map Z-score data for $tRNA^{phe}$ from *E. coli*; and secondary structure models of this RNA from **(b)** crystallography, **(c)** the *RNAstructure* algorithm without data, **(d)** calculations guided by one-dimensional SHAPE data, and **(e)** calculations guided by the two-dimensional mutate-and-map data. Cyan lines give model base pairs not present in crystallographic model; orange lines give crystallographic base pairs missed in each model. Bootstrap confidence estimates for each helix are given in red.

**Figure 3. Accurate secondary structure models for non-coding RNAs.** Mutate-and-map Z-score data and resulting secondary structure models for the P4-P6 domain of the *Tetrahymena* group I ribozyme [**(a)** & **(b)**], the 5S ribosomal RNA from *E. coli* [**(c)** & **(d)**], and the domain that binds cyclic di-guanosine monophosphate from the *V. cholerae* VC1722 riboswitch [in the presence of 10 μM ligand; **(e)** & **(f)**]. Cyan lines give model base pairs not present in crystallographic model; orange lines give crystallographic base pairs missed in each model. Bootstrap confidence estimates for each helix are given in red.

**Figure 4. Two states of a glycine-binding riboswitch.** Mutate-and-map Z-score data and resulting secondary structure models for the double-ligand-binding domain of the *F. nucleatum* glycine riboswitch with 10 mM glycine [**(a)** & **(b)**] and without glycine [**(c)** & **(d)**]. Bootstrap confidence estimates for each helix are given in red.

**Figure 5. Three-dimensional modeling from mutate-and-map data.** Models of the (ligand-bound) adenine riboswitch derived (a) from x-ray crystallography and (b) from guiding *de novo* modeling [by the Rosetta Fragment Assembly of RNA with Full Atom Refinement (FARFAR) algorithm] with secondary structure and tertiary contacts inferred from solution mutate-and-map data.

# Figure 1

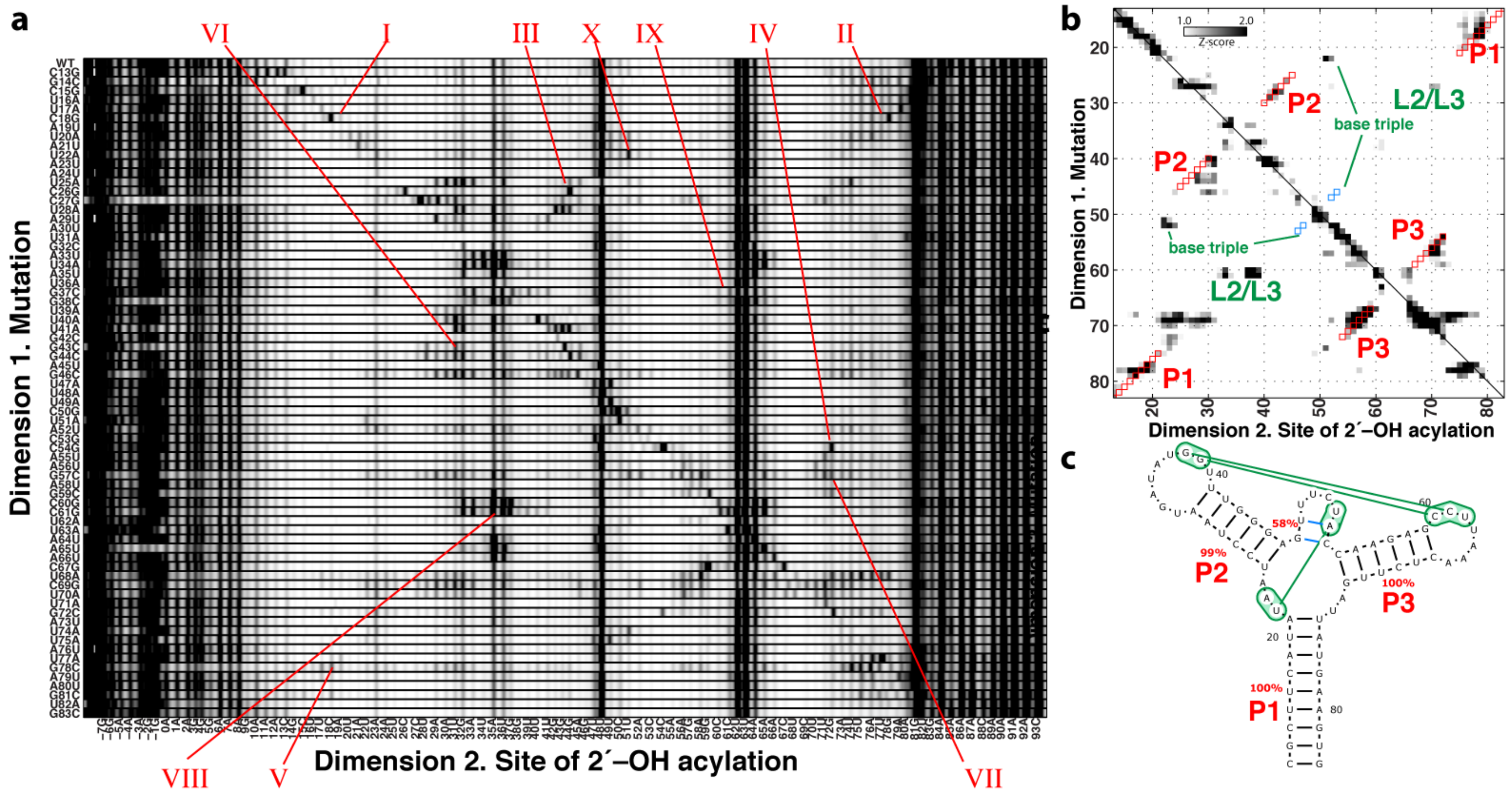

Figure 2

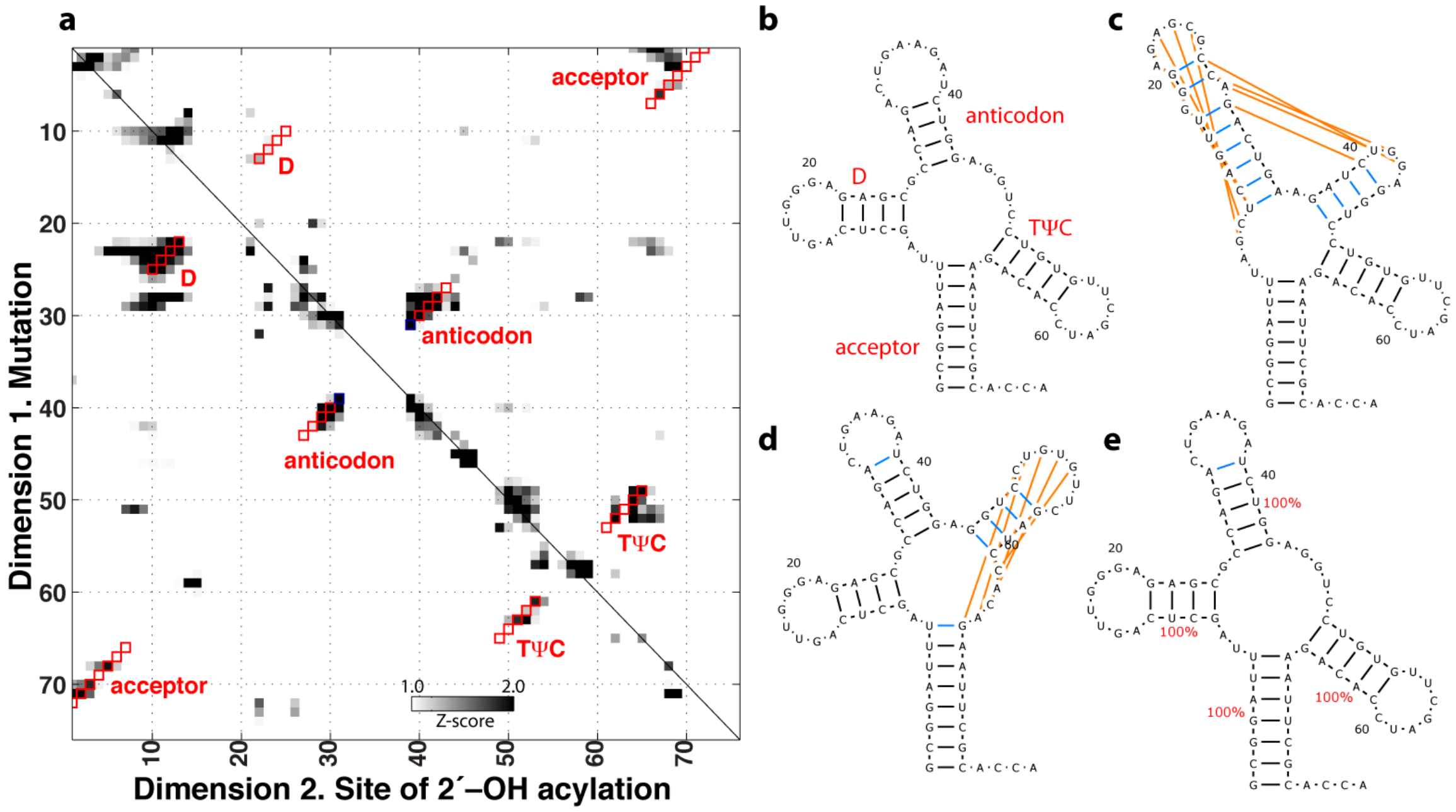

Figure 3

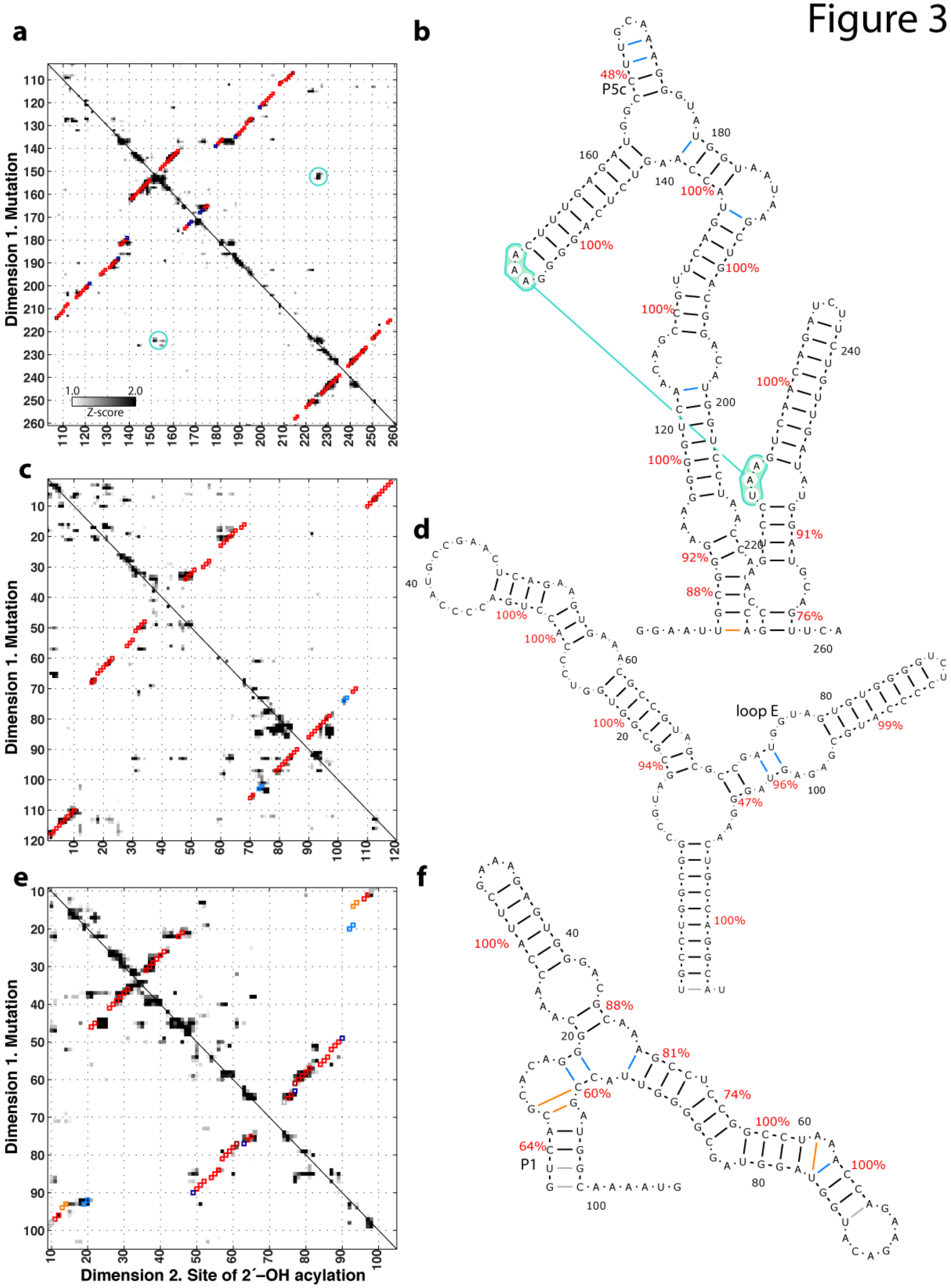

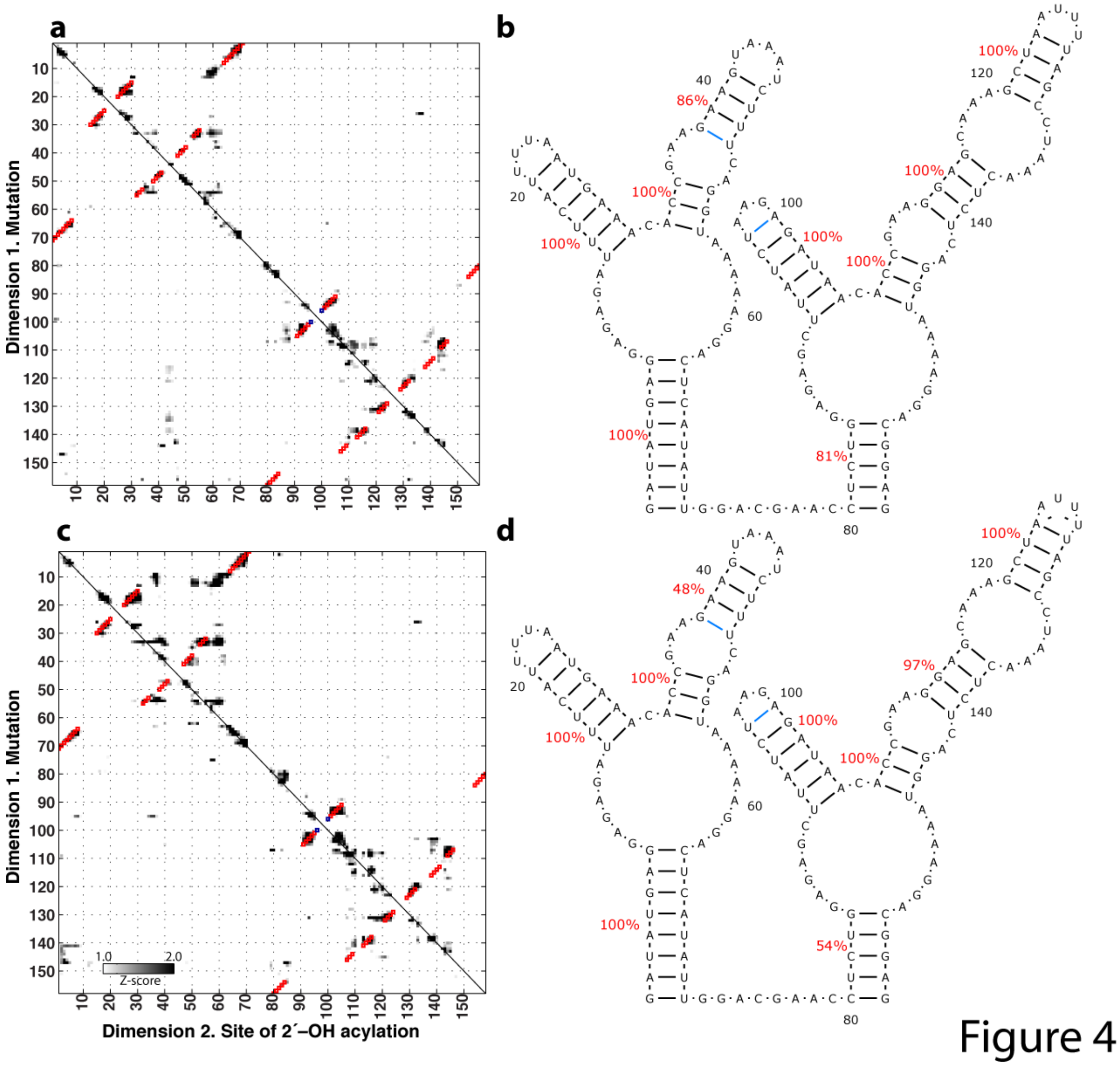

Figure 4

Figure 5

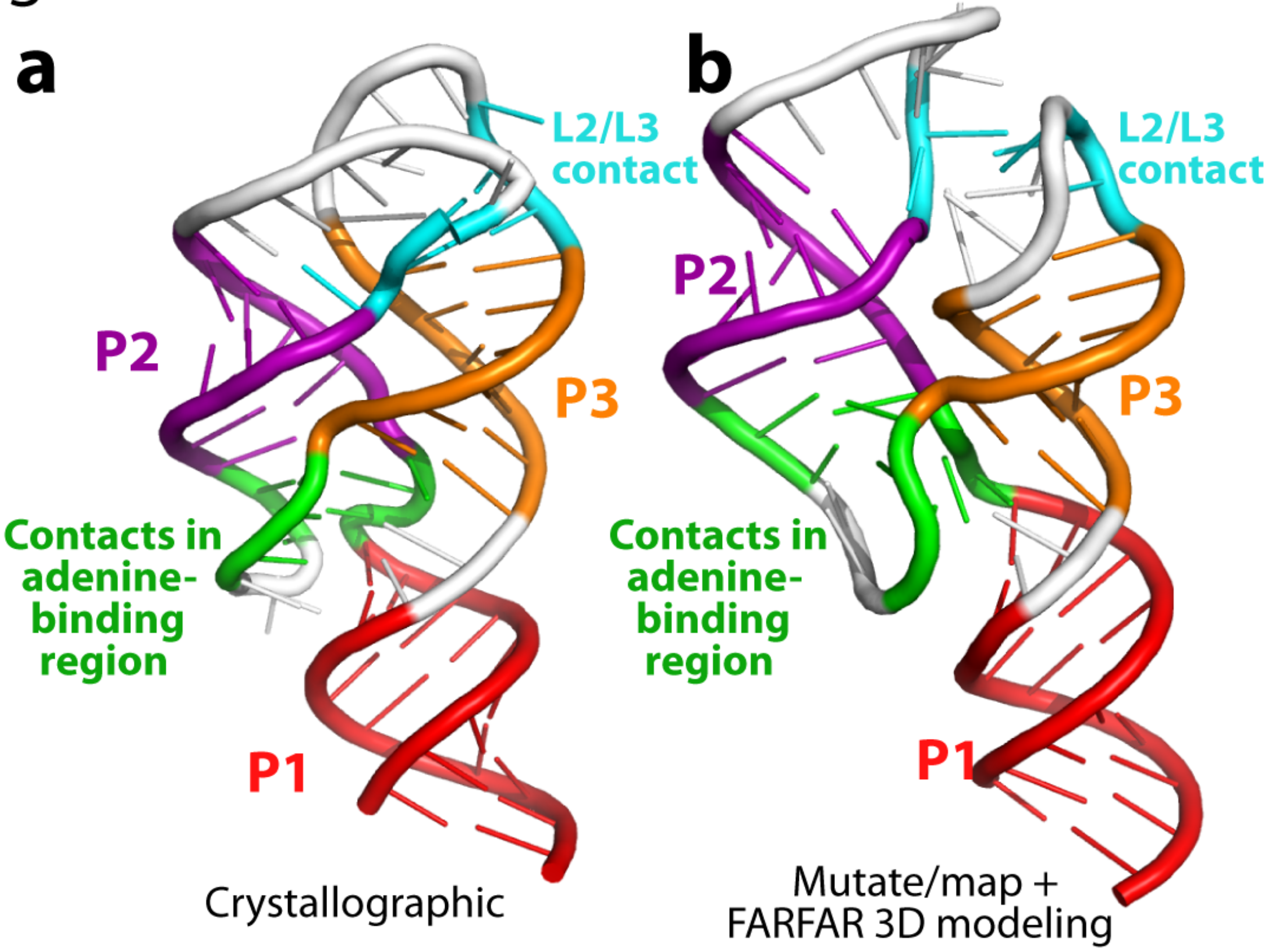

# **Supporting Information** for "A mutate-and-map strategy for non-coding RNAs: chemical inference of base pairs in two dimensions"

Wipapat Kladwang[1], Christopher C. VanLang[2], Pablo Cordero[3], and Rhiju Das[1,3,4]*

*Departments of Biochemistry[1] & Chemical Engineering[2], Program in Biomedical Informatics[3], and Department of Physics[4], Stanford University, Stanford CA 94305*

** To whom correspondence should be addressed: rhiju@stanford.edu. Phone: (650) 723-5976. Fax: (650) 723-6783.*

This document contains Supporting Methods, 2 Supporting Tables and 3 Supporting Figures.

## Supporting Methods

*Preparation of model RNAs*

The DNA templates for each RNA (Table S1) consisted of the 20-nucleotide T7 RNA polymerase promoter sequence (TTCTAATACGACTCACTATA) followed by the desired sequence. Double-stranded templates were prepared by PCR assembly of DNA oligomers up to 60 nucleotides in length (IDT, Integrated DNA Technologies, IA) with Phusion DNA polymerase (Finnzymes, MA). For each mutant, an automated MATLAB script was used to determine which primers required single-nucleotide changes and to generate 96-well plate spreadsheets for ordering and guiding pipetting for PCR assembly reactions.

Assembled DNA templates were purified in 96-well Greiner microplates with AMPure magnetic beads (Agencourt, Beckman Coulter, CA) following manufacturer's instructions. Sample concentrations were measured based on UV absorbance at 260 nm measured on Nanodrop 100 or 8000 spectrophotometers. Verification of template length was accomplished by electrophoresis of all samples and 10-bp and 20-bp ladder length standards (Fermentas, MD) in 4% agarose gels (containing 0.5 mg/mL ethidium bromide) and 1x TBE (100 mM Tris, 83 mM boric acid, 1 mM disodium EDTA).

In vitro RNA transcription reactions were carried out in 40 μL volumes with 10 pmols of DNA template; 20 units T7 RNA polymerase (New England Biolabs, MA); 40 mM Tris-HCl (pH 8.1); 25 mM $MgCl_2$; 2 mM spermidine; 1 mM each ATP, CTP, GTP, and UTP; 4% polyethylene glycol 1200; and 0.01% Triton-X-100. Reactions were incubated at 37 °C for 4 hours and monitored by electrophoresis of all samples along with 100–1000 nucleotide RNA length standards (RiboRuler, Fermentas, MD) in 4% denaturing agarose gels (1.1% formaldehyde; run in 1x TAE, 40 mM Tris, 20 mM acetic acid, 1 mM disodium EDTA), stained with SYBR Green II RNA gel stain (Invitrogen, CA) following manufacturer instructions. RNA samples were purified with MagMax magnetic beads (Ambion, TX), following manufacturer's instructions; and concentrations were measured by absorbance at 260 nm on Nanodrop 100 or 8000 spectrophotometers.

*SHAPE measurements*

Chemical modification reactions consisted of 1.2 pmols RNA in 20 μL with 50 mM Na-HEPES, pH 8.0, and 10 mM $MgCl_2$ and/or ligand at the desired concentration (see Table S1); and 5 μL of SHAPE modification reagent. The modification reagent was 24 mg/ml N-methyl isatoic anhydride freshly dissolved in anhydrous DMSO. The reactions were incubated at 24 °C for 15 to 60 minutes, with lower modification times for the longer RNAs to maintain overall modification rates less than 30%. In control reactions (for background measurements), 5 μL of deionized water was added instead of modification reagent, and incubated for the same time. Reactions were quenched with a premixed solution of 5 μL 0.5 M Na-MES, pH 6.0; 3 μL of 5 M NaCl, 1.5 μL of oligo-dT beads (poly(A) purist, Ambion, TX), and 0.25 μL of 0.5 mM 5´-rhodamine-green labeled primer (AAAAAAAAAAAAAAAAAAAAGTTGTTGTTGTTGTTTCTTT) complementary to the 3´ end of the MedLoop RNA [also used in our previous studies (*1, 2*)], and 0.05 μL of a 0.5 mM Alexa-555-labeled oligonucleotide (used to verify normalization). The reactions were purified by magnetic separation, rinsed with 40 μL of 70% ethanol twice, and allowed to air-dry for 10 minutes while remaining on a 96-post magnetic stand. The magnetic-bead mixtures were resuspended in 2.5 μL of deionized water.

The resulting mixtures of modified RNAs and primers bound to magnetic beads were reverse transcribed by the addition of a pre-mixed solution containing 0.2 μL of SuperScript III (Invitrogen, CA), 1.0 μL of 5x SuperScript First Strand buffer (Invitrogen, CA), 0.4 μL of 10 mM each dnTPs [dATP, dCTP, dTTP, and dITP (*3*)], 0.25 μL of 0.1 M DTT, and 0.65 μL water. The reactions (5 μL total) were incubated at 42 °C for 30 minutes. RNA was degraded by the addition of 5 μL of 0.4 M NaOH and incubation at 90 °C for 3 minutes. The solutions were neutralized by the addition of 5 μL of an acid quench (2 volumes 5 M NaCl, 2 volumes 2 M HCl, and 3 volumes of 3 M Na-acetate). Fluorescent DNA products were purified by magnetic bead separation, rinsed with 40 μL of 70% ethanol, and air dried for 5 minutes. The reverse transcription products, along with magnetic beads, were resuspended in 10 μL of a solution containing 0.125 mM Na-EDTA (pH 8.0) and a Texas-Red-labeled reference ladder (whose fluorescence is spectrally separated from the rhodamine-green-labeled products). The products were separated by capillary electrophoresis on an ABI 3100 or ABI 3700 DNA sequencer. Reference ladders for wild type RNAs were created using an analogous protocol without chemical modification and the addition of, e.g., 2´-3´-dideoxy-TTP in an amount equimolar to dTTP in the reverse transcriptase reaction.

*Data processing*

The HiTRACE software (*4, 5*) was used to analyze the electropherograms. Briefly, traces were aligned by automatically shifting and scaling the time coordinate, based on cross correlation of the Texas Red reference ladder co-loaded with all samples. Sequence assignments to bands, verified by comparison to sequencing ladders, permitted the automated peak-fitting of the traces to Gaussians. Data were normalized so that, within each mutant, the mean band intensity was unity for all nucleotides except the 20 nucleotides closest to the 5´ and 3´ ends. Individual replicate data sets, including aligned electropherograms and quantified band intensities, are being made publically available in the Stanford RNA Mapping Database (http://rmdb.stanford.edu).

For each data set, Z-scores were calculated as follows. Let the observed band intensities be $s_{ij}$ with $i = 1, 2, \ldots N$ indexing the band numbers, and $j = 1, 2, \ldots M$ indexing the nucleotides that were mutated. Then, the mean band intensities $\mu_i$ and standard deviations $\sigma_i$ were computed using their standarad definitions:

$$\mu_i = \frac{1}{N}\sum_{j=1}^{N} s_{ij}$$
$$\sigma_i = \left[\frac{1}{N}\sum_{j=1}^{N} (s_{ij} - \mu_i)^2\right]^{1/2} \qquad (1)$$

And the Z-scores were computed as:

$$Z_{ij} = [s_{ij} - \mu_i] / \sigma_i \qquad (2)$$

Only data with $Z_{ij} > 0.0$ and associated with bands with mean intensity $\mu_i$ less than a cutoff $\mu_i^{\mathrm{MAX}} = 0.8$ were kept, since the mutate-and-map approach seeks to identify site-specific release of nucleotides that are protected in the starting sequence and most variants. [Varying $\mu_i^{\mathrm{MAX}}$ from 0.5 to 1.0 gave indistinguishable results for models.] To avoid introducing additional noise, we did not correct for attenuation of longer reverse transcription products; because this effect should be similar for all mutants (and was observed to be such in the data), it scales $s_{ij}$, $\mu_i$, and $\sigma_{ij}$ identically (at a given nucleotide $i$) and did not affect the final $Z_{ij}$ scores in (2). Further, to again avoid introducing unnecessary noise, we did not explicitly subtract background measurements, as they should also subtract out of the Z-score expression (2). Nevertheless, control measurements for all RNAs without SHAPE modification were carried out. For a small number (<0.1%) of nucleotides in specific mutants, weak mutant-specific background bands were observed (likely due to sequence-specific reverse transcriptase stops). An analogous Z-score was carried out for these control background measurements; nucleotides with "background Z-scores" greater than 6.0 were identified as anomalous and set to zero in analyzing $Z_{ij}$ for mutate/map measurements. For data sets with more than one independent replicate (the *add* adenine-sensing riboswitch, the P4-P6 domain, and the *F. nucleatum* glycine-sensing riboswitch), $Z_{ij}$ values were averaged across the replicates. The analysis is available as a single MATLAB script *output_Zscore_from_rdat.m* within the HiTRACE package.

*Inference of contacts through sequence-independent clustering*

The Z-scores $Z_{ij}$ [see above, eq. (2)] define possible long-range contacts in each RNA. As in prior work (*1, 2*), mutate/map pairs with statistically strong signals ($Z_{ij} > Z_{\min}$; $Z_{\min}$= 1.0) were considered. The pair ($i$,$j$) was defined as neighboring any strong signals at ($i$–1, $j$), ($i$+1, $j$), ($i$, $j$–1), ($i$, $j$+1), or the symmetry partner ($j$,$i$); strong signals were then grouped by single-linkage clustering. Final selection of clusters used simple but stringent filters. Clusters with at least 8 pairs, involving at least three independent mutations, and including at least one pair of symmetry partners were taken as defining long-range interactions with strong support. For this selection, mutations involved in more than 5 such clusters were omitted as potentially being associated with cooperative, extended structural effects beyond the disruption of a single base pair, helix, or tertiary contact. The analysis is available as a single MATLAB script *cluster_z_scores.m* within the HiTRACE package.

*Three-dimensional modeling with Rosetta: command-lines and example files*

Three-dimensional models were acquired using the Fragment Assembly of RNA with Full Atom Refinement (FARFAR) methodology(*6*) in the Rosetta framework.

First, ideal A-form helices were created with the command line:

```
rna_helix.exe -database <path to database> -nstruct 1 -fasta
stem2_add.fasta -out:file:silent stem2_add.out
```

where the file `stem2_add.fasta` contains the sequence of the P2 helix, as determined by the mutate-and-map data:

```
>stem2_1y26.fasta
uccuaauugga
```

Then, for each RNA loop or junction motif that interconnects these helices, 4,000 models were created with FARFAR. For example, in the adenine riboswitch, two loops (L2 & L3) and the adenine-binding junction are the non-helical motif portions. The command line for building L2 onto the P2 helix is:

```
rna_denovo.<exe> -database  <path to database> -native
motif2_1y26_RNA.pdb -fasta motif2_add.fasta -params_file
motif2_add.params -nstruct 100 -out:file:silent motif2_add.out -cycles
5000 -mute all -close_loops -close_loops_after_each_move -minimize_rna
-close_loops -in:file:silent_struct_type rna -in:file:silent
stem2_add.out -chunk_res  1 2 3 4 5 6 16 17 18 19 20 21
```

Here, the optional “-native” flag, inputting the crystallographic structure for the motif, permits rmsd calculations but is not used in building the model.  The file `motif2_add.params` defines the P2 stem within this motif-building run:

```
STEM    PAIR 6 16 W W A PAIR 5 17 W W A PAIR 4 18 W W A PAIR 3 19 W W A
PAIR 2 20 W W A PAIR 1 21 W W A
OBLIGATE PAIR 1 21 W W A
```

Finally, the models of separately built motifs and helices are assembled through the FARNA Monte Carlo procedure:

```
rna_denovo.<exe> -database  <path to database> -native 1y26_RNA.pdb -
fasta add.fasta -in:file:silent_struct_type binary_rna  -cycles 10000 -
nstruct 200 -out:file:silent add_assemble.out -params_file
add_assemble.params -cst_file
add_mutate_map_threetertiarycontacts_assemble.cst -close_loops  -
in:file:silent  stem1_add.out stem2_add.out stem3_add.out
motif1_add.out motif2_add.out motif3_add.out -chunk_res  1 2 3 4 5 6 7
8 9 63 64 65 66 67 68 69 70 71 13 14 15 16 17 18 28 29 30 31 32 33 42
43 44 45 46 47 55 56 57 58 59 60 1 2 3 4 5 6 7 8 9 10 11 12 13 14 15 16
17 18 28 29 30 31 32 33 34 35 36 37 38 39 40 41 42 43 44 45 46 47 55 56
57 58 59 60 61 62 63 64 65 66 67 68 69 70 71 13 14 15 16 17 18 19 20 21
22 23 24 25 26 27 28 29 30 31 32 33 42 43 44 45 46 47 48 49 50 51 52 53
54 55 56 57 58 59 60
```

In the above command-line, the helix and loop definitions are given by `add_assemble.params`:

```
CUTPOINT_CLOSED  9 18 47
STEM  PAIR 1 71 W W A  PAIR 2 70 W W A  PAIR 3 69 W W A  PAIR 4 68 W W
A  PAIR 5 67 W W A  PAIR 6 66 W W A  PAIR 7 65 W W A  PAIR 8 64 W W A
PAIR 9 63 W W A
OBLIGATE PAIR 9 63 W W A
```

```
STEM  PAIR 13 33 W W A  PAIR 14 32 W W A  PAIR 15 31 W W A  PAIR 16 30
W W A  PAIR 17 29 W W A  PAIR 18 28 W W A
OBLIGATE PAIR 18 28 W W A

STEM  PAIR 42 60 W W A  PAIR 43 59 W W A  PAIR 44 58 W W A  PAIR 45 57
W W A  PAIR 46 56 W W A  PAIR 47 55 W W A
OBLIGATE PAIR 47 55 W W A
```

The constraint file `add_mutate_map_threetertiarycontacts_assemble.cst` encodes regions in tertiary contact (here including the short two-base-pair helix) inferred from the mutate-and-map data:

```
[ atompairs ]
N3 10 N3 39 FADE -100 10 2 -20.0
N3 10 N1 40 FADE -100 10 2 -20.0
N1 11 N3 39 FADE -100 10 2 -20.0
N1 11 N1 40 FADE -100 10 2 -20.0
N1 12 N1 40 FADE -100 10 2 -20.0
N3 10 N3 39 FADE -100 10 2 -20.0
N1 25 N3 49 FADE -100 10 2 -20.0
N1 25 N3 50 FADE -100 10 2 -20.0
N1 26 N3 48 FADE -100 10 2 -20.0
N1 26 N3 49 FADE -100 10 2 -20.0
N1 26 N3 50 FADE -100 10 2 -20.0
N3 27 N3 49 FADE -100 10 2 -20.0
N3 27 N3 50 FADE -100 10 2 -20.0
N3 35 N1 40 FADE -100 10 2 -40.0
N1 34 N3 41 FADE -100 10 2 -40.0
```

These constraints give a bonus of –20.0 kcal/mol if the specified atom pairs are within 8 Å; the function interpolates up to zero for distances by a cubic spline beyond 10.0 Å. Note that the Rosetta numbering here starts with 1 for the first nucleotide of the 71-nucleotide adenine binding domain, and is offset by 12 from the numbering in the crystal structure 1Y26. A total of 5000 models of the full-length RNA was generated, and the lowest-energy conformation was taken as the final model. (For the adenine riboswitch, the next nine lowest energy models were within 2 Å RMSD of this model, indicating convergence.)

## Table S1. Benchmark for the mutate-and-map strategy for noncoding RNA base pair inference.

| RNA, source | Solution conditions[a] | Off–set[b] | PDB[c] | Sequence & Secondary Structure[d] |
|---|---|---|---|---|
| Adenine riboswitch, *V. vulnificus* (*add*) | Standard + 5 mM adenine | –8 | **1Y26**<br>1Y27<br>2G9C<br>3GO2<br>… | ggaaaggaaagggaaagaaaCGCUUCAUAUAAUCCUAAUGAUAUGGUUUGGG<br>AGUUUCUACCAAGAGCCUUAAACUCUUGAUUAUGAAGUGaaaacaaaacaaa<br>gaaacaacaacaacaac<br>....................(((((((((...((((((.........)))))<br>)........((((((.......))))))..))))))))).............<br>................. |
| tRNA[phe], *E. coli* | Standard | –15 | **1L0U**<br>1EHZ | ggaacaaacaaaacaGCGGAUUUAGCUCAGUUGGGAGAGCGCCAGACUGAAG<br>AUCUGGAGGUCCUGUGUUCGAUCCACAGAAUUCGCACCAaaaccaaagaaac<br>aacaacaacaac<br>...............(((((((..((((........)))).((((.......<br>..)))).....(((((.......))))))))))))).................<br>............ |
| P4–P6 domain, *Tetrahymena* ribozyme | Standard, 30% methyl-pentanediol[e] | 89 | **1GID**<br>1L8V<br>1HR2<br>2R8S | ggccaaaacaacgGAAUUGCGGGAAAGGGGUCAACAGCCGUUCAGUACCAAG<br>UCUCAGGGGAAACUUUGAGAUGGCCUUGCAAAGGGUAUGGUAAUAAGCUGAC<br>GGACAUGGUCCUAACCACGCAGCCAAGUCCUAAGUCAACAGAUCUUCUGUUG<br>AUAUGGAUGCAGUUCAaaaccaaaccaaagaaacaacaacaacaac<br>.................((((((...((((((.....(((.((((.(((..(<br>((((((((....)))))))))..((.......))....)))......)))))<br>))....))))))..)).))))((...((((...(((((((((...)))))))<br>))..))))...)).................................. |
| 5S rRNA, *E. coli* | Standard | –20 | **3OFC**<br>3OAS<br>3ORB<br>2WWQ<br>… | ggaaaggaaagggaaagaaaUGCCUGGCGGCCGUAGCGCGGUGGUCCCACCU<br>GACCCCAUGCCGAACUCAGAAGUGAAACGCCGUAGCGCCGAUGGUAGUGUGG<br>GGUCUCCCCAUGCGAGAGUAGGGAACUGCCAGGCAUaaaacaaaacaaagaa<br>acaacaacaacaac<br>.....................(((((((((.....((((((((....(((((<br>((.............))))..)))...)))))).)).((.......((((((<br>((...))))))))).......))...)))))))))..................<br>.............. |
| Cyclic di-GMP riboswitch, *V. cholerae* (*VC1722*) | Standard + 10 μM cyclic diguanosine monophosphate | 0 | **3MXH**<br>3IWN<br>3MUV<br>3MUT<br>… | ggaaaauGUCACGCACAGGGCAAACCAUUCGAAAGAGUGGGACGCAAAGCC<br>UCCGGCCUAAACCAGAAGACAUGGUAGGUAGCGGGGUUACCGAUGGCAAAAU<br>Gcauacaaaccaaagaaacaacaacaacaac<br>..........((((......((...((((((....))))))...))...(((<br>.((((((((..((.........)))))))..))))))...)).)).......<br>............................... |
| Glycine riboswitch, *F. nucleatum* | Standard + 10 mM glycine | –10 | **3P49** | ggacagagagGAUAUGAGGAGAGAUUUCAUUUUAAUGAAACACCGAAGAAGU<br>AAAUCUUUCAGGUAAAAAGGACUCAUAUUGGACGAACCUCUGGAGAGCUUAU<br>CUAAGAGAUAACACCGAAGGAGCAAAGCUAAUUUUAGCCUAAACUCUCAGGU<br>AAAAGGACGGAGaaaacacaacaaagaaacaacaacaacaac<br>..........((((((((......((((((....)))))).(((...((((.<br>.....))))..))).......))))))))........(((((......((((<br>(.....))))).(((...((((....((((....)))).....))))..)))<br>.......))))).............................. |

[a] Standard conditions are: 10 mM $MgCl_2$, 50 mM Na–HEPES, pH 8.0 at 24 °C.

[b] Number added to sequence index to yield numbering used in previous biophysical studies, and in Fig. 1 of the main text.

[c] The first listed PDB ID was the source of the assumed crystallographic secondary structure; other listed IDs contain sequence variants, different complexes, or different crystallographic space groups and confirm this structure.

[d] In the sequence, lowercase symbols denote 5´and 3´ buffer sequences, including primer binding site (last 20 nucleotides). In all cases, designs were checked in RNAstructure and ViennaRNA to give negligible base pairing between added sequences and target domain. Structure is given in dot–bracket notation, and here denotes base pairs for which there is crystallographic evidence. A long-range two-base-pair helix [25–50, 26–49] in the adenine riboswitch is involved in an extensive non-canonical loop-loop interaction and is not included.

[e] 2-methyl-2,4,-pentanediol (MPD) was included due to reports that its presence in crystallization buffer can change SHAPE reactivity of the P4-P6 RNA (Vicens et al. (2007) RNA 13, 536–48). Mutate/map measurements without MPD gave different reactivities in the P5abc region and ambiguous modeling results in the P5c region; the crystallographic helix or a helix with single-nucleotide register shift was observed in models from different bootstrap replicates.

**Table S2. Base-pair-resolution analysis of the helices recovered by the mutate-and-map method.**

| RNA | Crystallographic | Correctly recovered | Missed[a] | | | Extra base pairs[a] | | |
|---|---|---|---|---|---|---|---|---|
| | | | A-U | G-U | G-C | A-U | G-U | G-C |
| Adenine ribosw.[b] | 21 | 21 | 0 | 0 | 0 | 0 | 0 | 0 |
| tRNA$^{phe}$ | 20 | 20 | 0 | 0 | 0 | 1 | 0 | 0 |
| P4-P6 RNA | 48 | 47 | 1 | 0 | 0 | 4 | 1 | 0 |
| 5S rRNA | 34 | 34 | 0 | 0 | 0 | 0 | 0 | 0 |
| c-di-GMP ribosw.[b] | 25 | 23 | 0 | 0 | 0 | 2 | 0 | 0 |
| Glycine ribosw.[b] | 40 | 40 | 0 | 0 | 0 | 1 | 2 | 0 |
| *Total* | 188 | 185 | 1 | 0 | 0 | 8 | 3 | 0 |

[a] Number of missed or extra base-pairs within helices that match crystallographic helices. (The only crystallographic helix not recovered in the mutate-and-map models is a short stem with two G-C base pairs in the cyclic di-GMP riboswitch.)

[b] Ligand-binding riboswitches were probed in the presence of small-molecule partners (5 mM adenine, 10 μM cyclic di-guanosine-monophosphate, or 10 mM glycine). All experiments were carried out with 10 mM $MgCl_2$, 50 mM Na-HEPES, pH 8.0.

**Figure S1.** Mutate-and-map analysis of a partially ordered state of the adenine riboswitch. (a) Mutate-and-map data (Z-scores) are given in gray-scale for the adenine-binding domain from the *add* riboswitch, *V. vulnificus*, without adenine present. Red squares mark crystallographic secondary structure of the RNA in its adenine-bound form. (b) Dominant secondary structure for the ligand-free adenine riboswitch, inferred from mutate-and-map data, is not distinguishable from the ligand-bound form (see Main Text Fig. 1c). Cyan lines mark an 'extra' helix that is also seen in the ligand-bound state; the helix corresponds to neighboring Watson-Crick base pair and non-Watson-Crick base pair seen in the crystallographic ligand-bound model. Bootstrap confidence estimates for each helix are given in red.

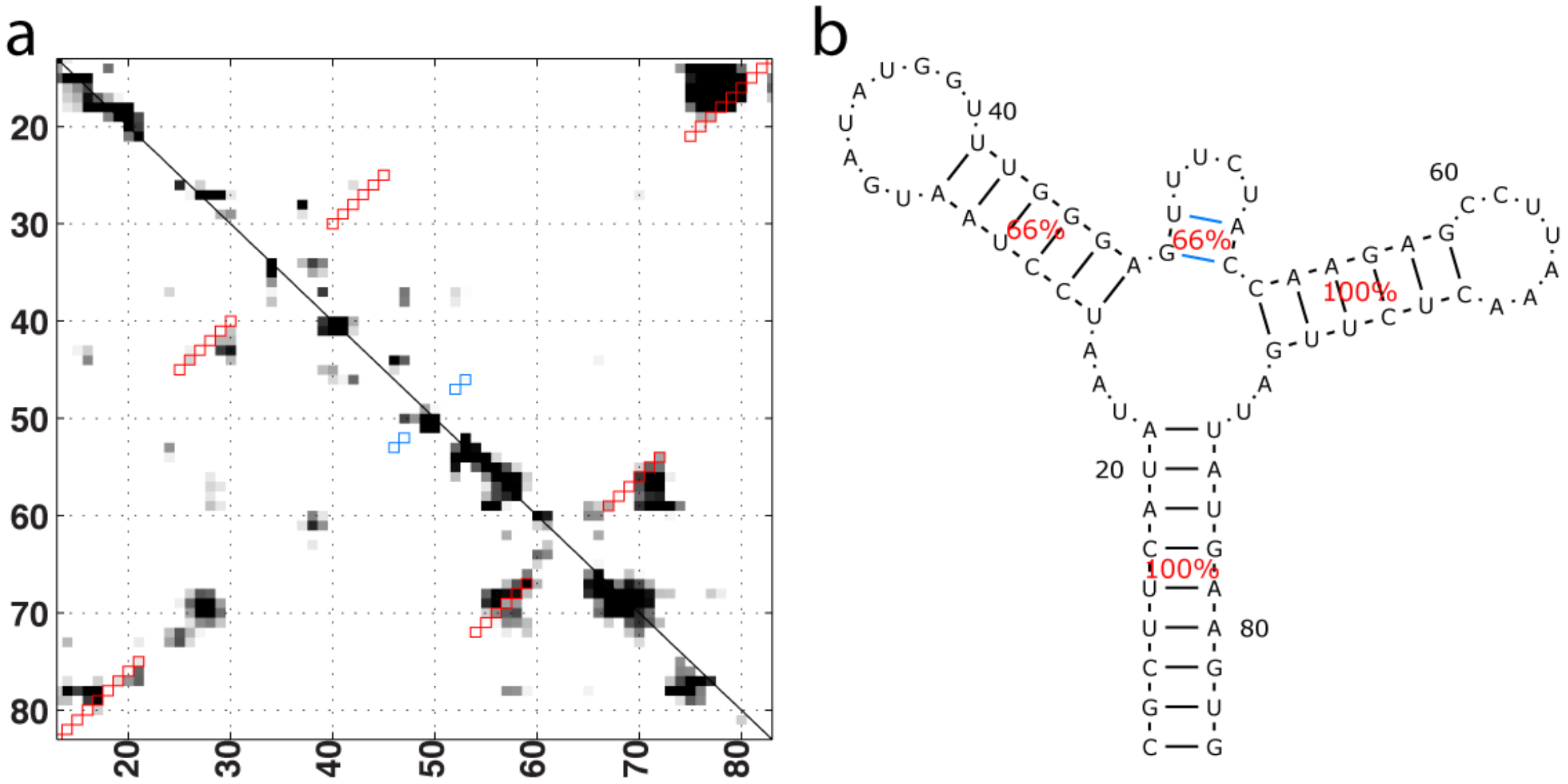

**Figure S2.** Mutate-and-map analysis of a partially ordered state of a cyclic di-guanosine monophosphate (c-di-GMP) riboswitch. (a) Mutate-and-map data (Z-scores) are given in gray-scale for the c-di-GMP-binding domain from the *VC1722* riboswitch, *V. cholerae*, without c-di-GMP present. Red squares mark crystallographic secondary structure of the RNA in its c-di-GMP-bound form. (b) Secondary structure models for this ligand-free c-di-GMP riboswitch, inferred from mutate-and-map data, is different from models for the ligand-bound state near the P1 stem and c-di-GMP binding region. Cyan lines are mutate-and-map base pairs not present in (ligand-bound) crystal structure, and orange lines are crystallographic base pairs not present in the mutate-and-map model. Bootstrap confidence estimates for each helix are given in red.

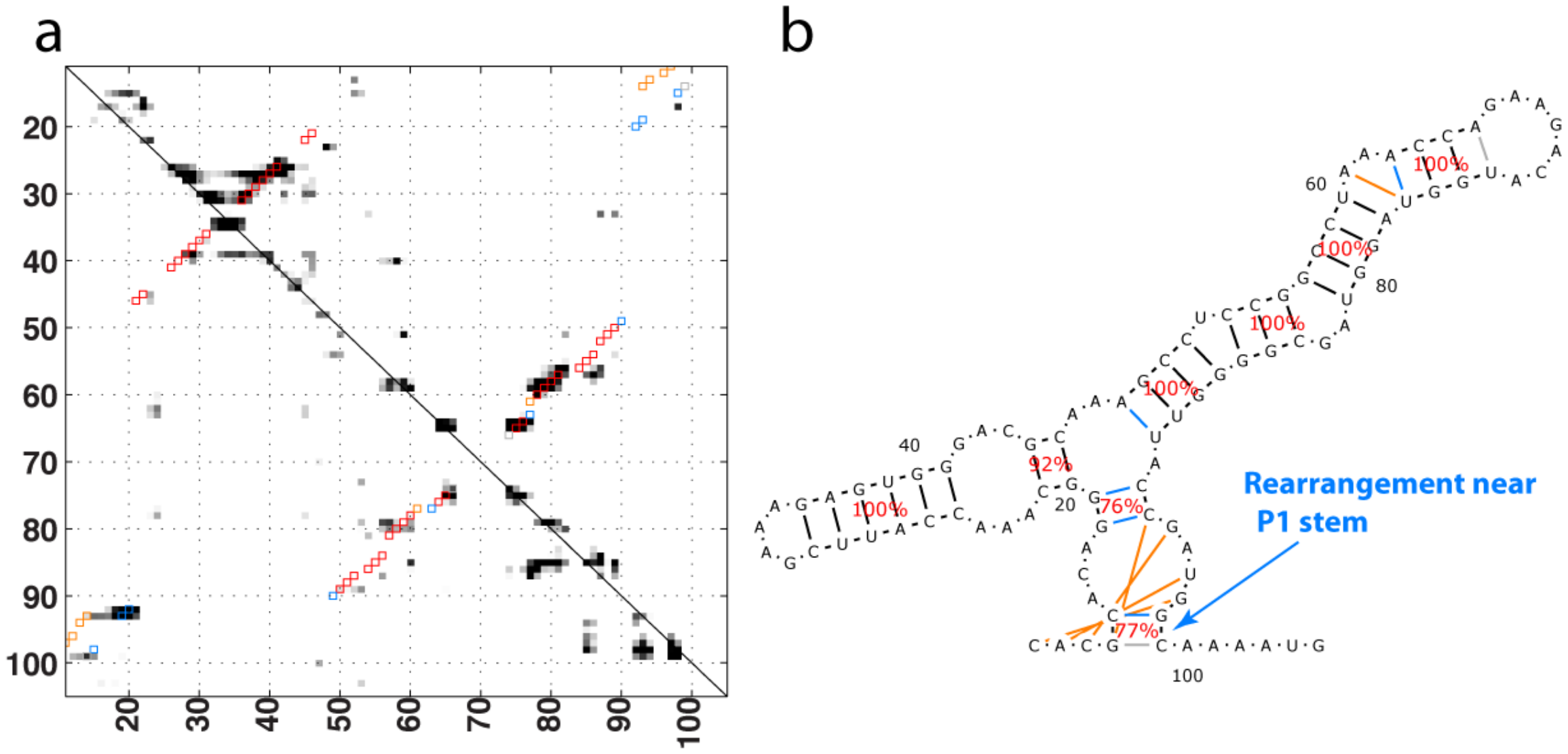

**Figure S3. Accurate inference of contacting regions in structured non-coding RNAs through sequence-independent analysis of mutate-and-map data**. Cluster analysis of *Z*-scores, using filters for signal strength, number of independent mutants, and symmetry of features (see Methods); final clusters are shown in different, randomly chosen colors. Base pairs from crystallographic secondary structure are marked as black symbols. RNAs are (a) adenine riboswitch, (b) tRNAphe, (c) P4-P6 RNA, (d) 5S rRNA, (e) c-di-GMP riboswitch, and (f) glycine riboswitch. Riboswitch data were collected with ligands present.

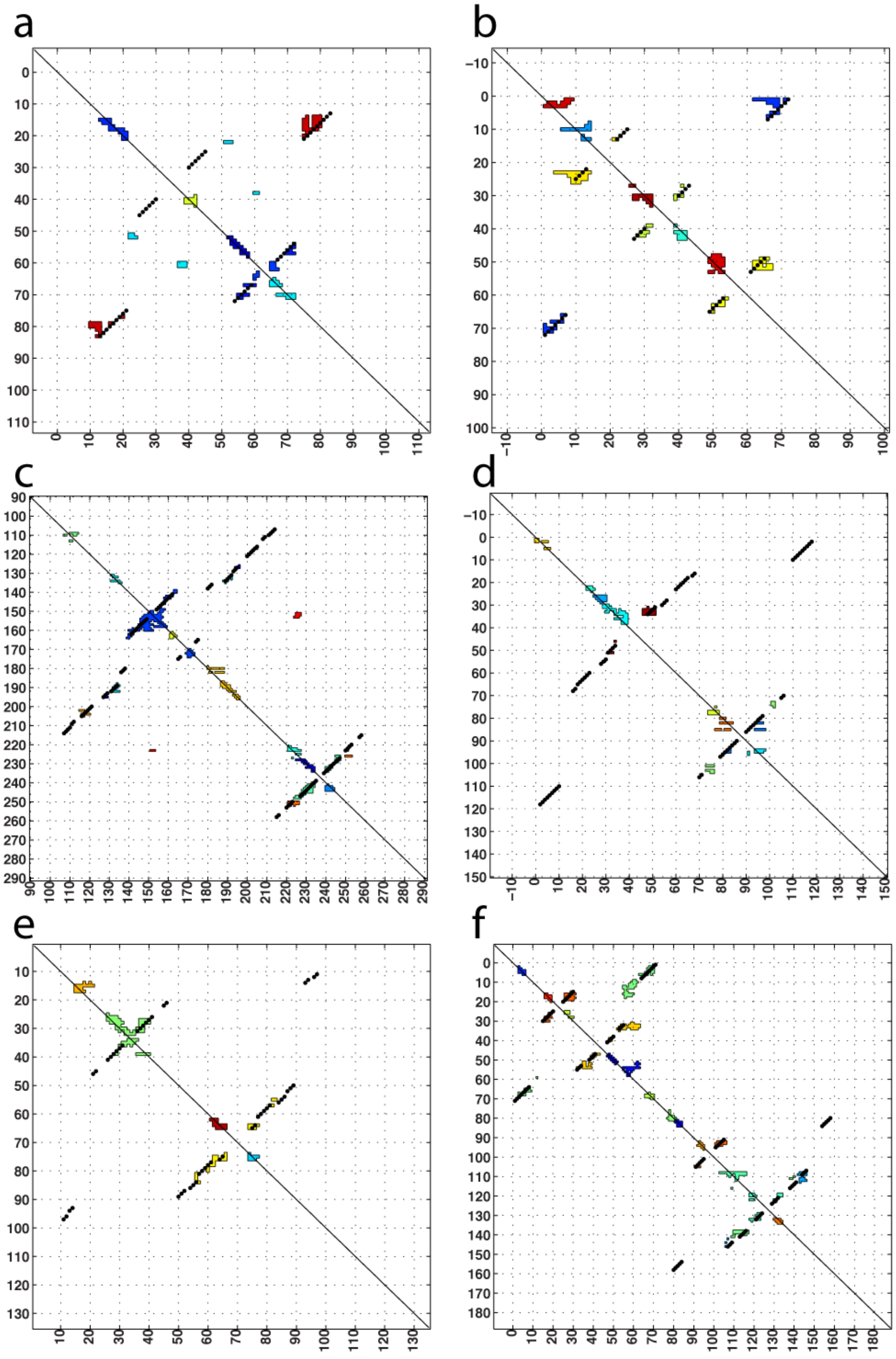